\documentclass[12pt]{article}
\pdfoutput = 1
\usepackage{graphicx} 
\usepackage{array}
\usepackage{authblk}

\usepackage{placeins}
\usepackage{hyperref}
\begin{document}

\title{Efficient determination of bespoke optically active nanoparticle distributions}

\author[1,2]{Phillip Manley\thanks{phillip.manley@helmholtz-berlin.de}}
\author[1]{Min Song}
\author[2]{Sven Burger}
\author[1,3]{Martina Schmid}

\affil[1]{Helmholtz-Zentrum-Berlin f\"{u}r Material und Energie, Kekul\'{e}str. 5, 12489 Berlin, Germany}
\affil[2]{Computational Nanooptics, Zuse Institute Berlin, Takustr. 7, 14195 Berlin, Germany.}
\affil[3]{University of Duisburg-Essen and CENIDE, Lotharstr. 1, 47057 Duisburg, Germany}
\maketitle

This paper was published in the Journal of Optics \textbf{20}(8) p.085003 (2018) doi: \href{dx.doi.org/10.1088/2040-8986/aad114}{10.1088/2040-8986/aad114} and is made available as an electronic preprint with permission from the Institute of Physics.

\begin{abstract}
	We provide a computational method for quickly determining the correct distribution of optically active nanoparticles for a desired response. This is achieved by simulating the optical response of single nanoparticles and performing a statistical averaging over different sizes. We find good agreement between experiment and theory for the transmission, reflectance and absorption of both an ordered and disordered array. By repeating the simulation for different particle distributions, we show that the method is capable of accurately predicting the correct nanoparticle distribution for a desired optical response. We provide a referential graph for predicting the optical response of different Ag nanoparticle distributions on a glass substrate, which can be extended to other substrate and particle materials, and particle shapes and sizes.
\end{abstract}


\section{Introduction}
\begin{figure}[h]
\centering
  \includegraphics[width=0.5\textwidth]{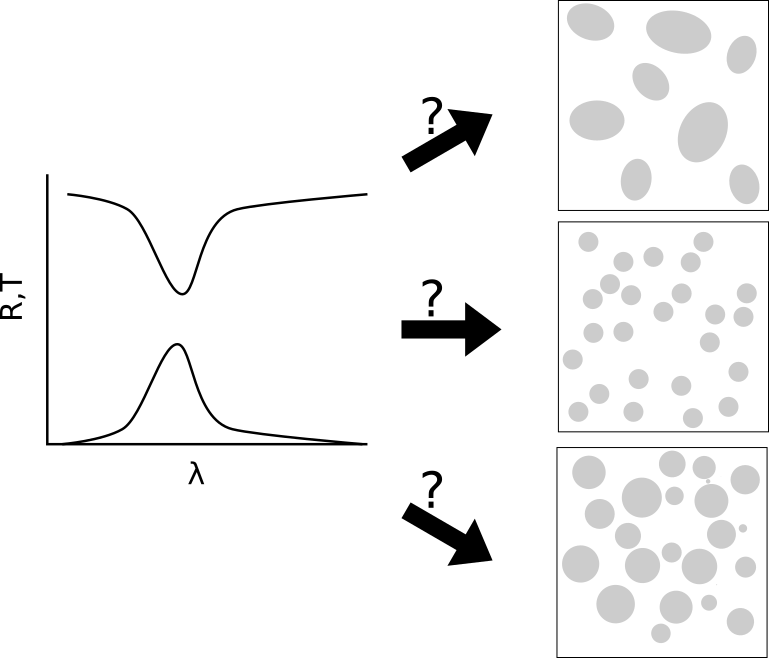}
  \caption{The often encountered design problem: given the desired optical response, which nanoparticle distribution is needed?}
  \label{fig:Inverse_Problem}
\end{figure}

Nanoparticles which have a strong interaction with light, i.e optically active nanoparticles, have been shown to be a key future technology for the manipulation of light on the nanoscale \cite{Abalde2010,Atwater2010,Evanoff2005,Hylton2013,Kelly2003,Kuznetsov2012,Huang2011,Pankhurst2003,Leung2014,Shin2015,Shao2014}. For realistic applications the response of many particles in an array is typically considered. The array can be either ordered meaning that a periodically repeated unit cell can be defined or disordered when no such unit cell can be defined. It is generally possible to accurately compute the optical response of a given array, but it may be computationally expensive depending on the array properties. However, in general the more important question to be answered is not, \textit{what is the optical response of an array?}, but rather, \textit{which kind of array provides the desired optical response?} i.e. the design problem as shown in figure \ref{fig:Inverse_Problem}. This is closely related to the field of inverse problems. Although there is a rich mathematical literature on solving inverse problems\cite{Colton1996,Ratnajeevan1993,Sarvas1987}, the technique most commonly employed to solve a design problem is to compute the output for a range of input parameters which requires solving the forward problem many times. This automatically provides information about which regions of input parameters give close to optimal solutions (which can be just as critical as finding the optimum solution itself). This is the approach which will be followed in this paper.

For large parameter scans to be feasible, it requires a fast method for solving the forward problem. This paper focuses on a method for rapidly calculating the properties of optically active nanoparticle distributions (the forward problem). In order to model nanoparticles appropriately, the ratio of nanoparticle size to the wavelength of light should be considered. For nanoparticles that are small compared to the wavelength, an effective medium approach may be appropriate \cite{Foldy1945,Siraji2017}.

If treating the nanoparticles as distinct objects, a dipole approach can be used \cite{Cortes2015,Zou2005,Schmid2011}. For nanoparticles at optical wavelengths, the particle size may be comparable to the wavelength, necessitating other models. For single spherical particles an analytical solution for the interaction with light has been known since 1908 in the multipole expansion commonly referred to as Mie theory \cite{Mie1908}. Multiple multipoles can be coupled together in order to model a complete array \cite{Xu1995}. If combined with numerical techniques, the scattered field from an arbitrarily shaped isolated scatterer can be decomposed into dipoles \cite{Waterman1965} or multipoles \cite{Grahn2012,David2012} which can then couple to each other. Recent approaches aim to tackle dielectric particles on conductive substrates \cite{Li2018}.



Nanoparticle arrays may also be handled using numerical techniques, such as the discrete dipole approximation \cite{Draine1994,Kim2015},  Fourier model method \cite{Meisenheimer2014,Haitong2018}, finite differences \cite{Hohenau2007,Mokkapati2009,Yang2015}, and the finite element method (FEM) \cite{McMahon2009,Yin2015,Burger2010}. Numerical techniques are capable of modelling arbitrary geometries and therefore can include the effect of a substrate on an array of nanoparticles. This comes at a greater computational cost in comparison to analytical techniques.

For ordered arrangements of particles the computational cost may remain acceptable through the use of periodic boundary conditions. However for disordered arrays, the large amount of particles needed for a simulation may move the computational cost to unacceptably high levels. In this case we present a method whereby individual particle responses can be used to obtain the full array response. This will involve neglecting inter-particle coupling which is valid for low densities of nanoparticles with random positions.

Firstly the proposed method of statistical averaging will be set out in the Theory section. The numerical method used for obtaining single particle responses, which is independent of the proposed method of statistical averaging, and the experimental methods used for obtaining the nanoparticle samples presented are given in the Methods section. The results of the proposed method are then discussed in detail, in particular two comparisons of theory to experiment are presented. Additionally, the specificity of the method is demonstrated by performing a parameter scan over many input parameters. Finally a reference of the optical response for Ag nanoparticle distributions on glass with various distribution parameters is given which can be used as a guideline for finding the correct nanoparticle distribution for a given application.

\section{Theory}
The response of a system of optically active nanoparticles can be very complicated and therefore computationally demanding to calculate. Analytical methods have been proposed for simulating arrays of particles, but are generally limited to using simple geometrical shapes for the particles such as spheres. More flexible numerical techniques such as the FEM can simulate particles with a very high degree of accuracy at the cost of computational effort. In order to remain general with regards to the type of particle shape the FEM is employed for all presented simulations, however it should be noted that the proposed method is independent of a particular numerical method used. As shall be seen, the only requirement for the numerical method is that it can deliver the response of single particles.

The simplest case computationally is that of periodically repeating system of nanoparticles. Consider an array of nanoparticles in the $x-y$ plane, positioned at an interface at $z$=0. By identifying the periodically repeating unit cell and employing Floquet boundary conditions in the x-y plane and transparent boundary conditions in the $z$ direction, the optical response can be calculated as long as the period of the unit cell is sufficiently small. Transparent boundary conditions are also commonly known as perfectly matched layer (PML) boundaries \cite{Berenger2007}.

\begin{figure}[h]
\centering
  \includegraphics[width=0.48\textwidth]{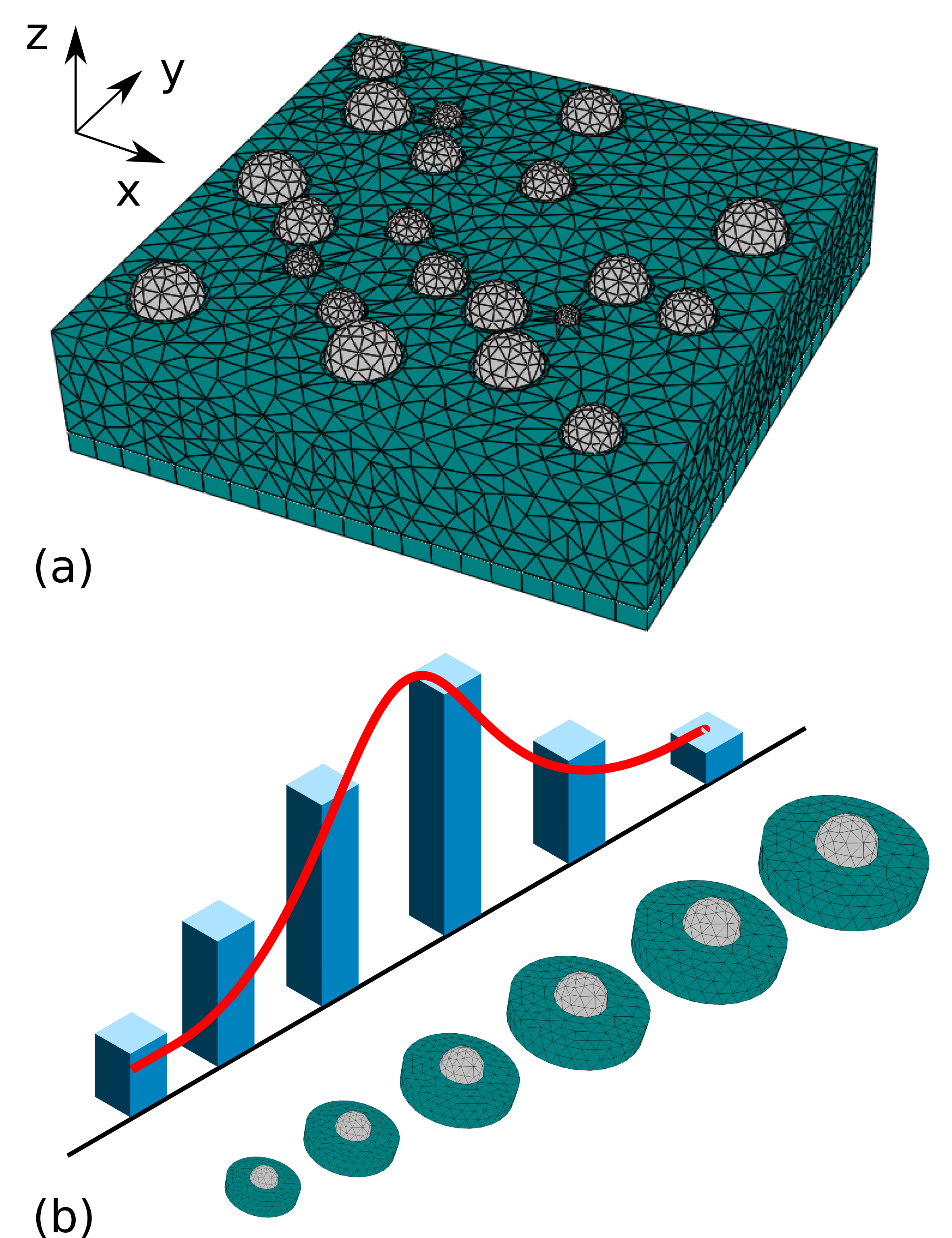}
  \caption{(a) The FEM mesh used to simulate the disordered array found in figure \ref{fig:Disordered}, upper layer of air has been hidden for visibility. (b) Breakdown of part (a) into individual particles to be simulated separately weighted by their statistical probability.}
  \label{fig:Grids}
\end{figure}

A much more difficult problem is that of a disordered array of nanoparticles, such as the one shown in figure \ref{fig:Grids}. In order to approximate the disorder sufficiently, a large number of nanoparticles must be contained inside the computational domain (CD) which typically means using a correspondingly large CD. The choice of boundary conditions in the x-y plane can be either Floquet or PML boundary conditions, either case will give approximate results, where the accuracy of the approximation will increase as the CD size increases, due to the diminishing contribution of edge effects. Both of these factors point towards the necessity of large CDs which means long computation times or in the worst case intractable problems. For an optimisation process which requires determining the optical response for many different configurations this approach is clearly unsuitable.

Therefore, a two-step method is proposed whereby the response of single nanoparticles is simulated separately in the first step, followed by combining the different responses to obtain the response of the entire array in the second step.

The validity of this method is based on the following assumptions
\begin{enumerate}
\item	Nanoparticles are sufficiently separated such that there are no near field interactions between them.
\item	Nanoparticles are randomly distributed, meaning that there is no fixed phase correlation between them.
\end{enumerate}
Under these conditions we can consider the nanoparticles to be independent and uncorrelated. Thus the average response of an array of particles will be given by the expectation value of the individual particle responses weighted by an appropriate probability density function. As an example, the reflectance of an array of particles can be calculated by,

\begin{eqnarray}
<R(\lambda)> = \int_{0}^{\infty}{R(\lambda,r)\rho(r)dr}, \\
\rho(r) = \frac{1}{\sigma\sqrt{2\pi}} exp\left[\frac{(r-\mu)^{2}}{2\sigma^{2}}\right].
\label{eq:NormalDist}
\end{eqnarray}

Where R($\lambda$,$r$) is the wavelength and particle radius dependent reflectance, and $\rho$($r$) is the probability density function describing the probability to find a particle with radius $r$. In this case the probability density function used was a normal distribution with mean $\mu$ and standard deviation $\sigma$, however the method is independent of the particular distribution used. Analogous equations can be used to obtain the transmission and absorption responses which fully characterises the optical response of the particles. The integral is then solved numerically since there is in general no analytical expression for the reflectance, transmission and absorption as a function of particle size and wavelength for particles at an interface.

To obtain the wavelength dependent response of the individual particles, each particle is simulated using the FEM and PML boundary conditions in all dimensions. Figure \ref{fig:Grids} displays the concept of reducing the full multi-particle FEM problem down to many individual isolated particle problems. For each particle the average coverage of the array is taken to be equal to the coverage of the particle in the simulation, i.e. the ratio of particle- and domain-cross-sectional area is equal to the overall particle coverage. It should be noted that if the coverage becomes too large, the total extinction efficiency may become larger than the computational domain cross sectional area. This would mean that more light could interact with the nanoparticle than was incident to the domain which would invalidate the approach. This violates the first assumption that there is no near field interaction between the particles. In order to fulfill this condition it is suggested to keep the inter-particle distance high enough to ensure that extinction efficiency multiplied by the coverage is less than unity for all wavelengths.

The results here are presented for nanoparticle arrays on a substrate. That is due to the difficulty previously outlined in handling this particular geometry. The case of well separated particles suspended in bulk is simpler due to being able to consider the surroundings of the particles as homogeneous. In that case the method presented here can be used for low density suspensions of particles. To achieve this the single particle responses, which we calculate numerically, should be calculated for isolated particles in a homogeneous medium. For the case of spherical particles, the single particle responses can be determined using Mie theory. If inter-particle interactions are non-negligible then a multipole decomposition of the scattered field of each nanoparticle (calculated via Mie theory for spheres or numerically for arbitrary geometries) can be used to determine the coupled response. This is not possible for the case presented here as the multipole decomposition is not consistent with a non-homogeneous surrounding such as a substrate.

\section{Methods}
The proposed simulation procedure is independent of the particular method used to produce the nanoparticles. Therefore different methods for preparing nanoparticles such as chemical reduction \cite{Wang2005,Choi2007,Giallongo2013}, electron beam lithography \cite{Corbierre2005} and laser ablation \cite{Haustrup2011} can all be analysed. In order to produce the samples presented in this work, a simple technique based on thermal annealing was used \cite{Morawiec2013,Santbergen2012,Thouti2013}. In order to prepare the disordered array, a 20 nm thin film of Ag is first deposited on a glass substrate via evaporation under vacuum conditions. The Ag film is then annealed at a temperature of 500 $^{\circ}$C for 20 minutes in air resulting in the formation of nanoparticles. In order to create the ordered array the same technique was employed with the additionally step of polystyrene (PS) spheres being deposited before the Ag film which act as a mask for the Ag film. The PS latex solution is mixed with a solution of 1\% styrene in ethanol by a 1:1 volume ratio. A glass substrate is then submerged in water in a Petri dish. The PS spheres are placed on the water surface using a pipette resulting in a hexagonally close packed monolayer. The water is then sucked out which transfers the PS monolayer to the glass substrate surface. A 60 nm thickness Ag film is then evaporated onto the PS sphere mask. The resulting structure is submerged in toluene and cleaned in an ultrasonic bath for 10 minutes in order to remove the PS spheres, leaving a honeycomb arrangement of triangular Ag nanoparticles. Finally thermal annealing at 500 $^{\circ}$C for 15 minutes changes the nanoparticle shape from triangular to spherical while preserving their ordered positions. This method is typically referred to as nanosphere lithography \cite{Kosiorek2004}.

The images used for the statistical analysis of particle distributions were taken using a scanning electron microscope (SEM). These images were then processed using the ImageJ software in order to obtain the radius of each particle as well as the overall coverage ($\rho$) of particles \cite{ImageJ}. The transmittance and reflectance of the particle arrays were measured using a PerkinElmer lambda 950 spectrophotometer with an integrating sphere. Due to the integrating sphere, the reflectance and transmittance include both the specular and scattered parts. The absorption is determined via energy conservation considerations.

In order to simulate both particle arrays and single particles, the FEM package JCMsuite is employed \cite{Pomplun2007}. The FEM is capable of simulating arbitrary particle shapes, however, in the samples presented here, the particles showed a high degree of rotational symmetry. In this case to increase the computational speed, rotationally symmetric particles were assumed. This allows a cylindrical coordinate system to be used, where the FEM problem is solved in the (r,z) plane while the $\theta$ dependence is determined by a Fourier expansion. The reflectance and transmittance are calculated using the surface integral of the Poynting flux through the interfaces to the upper and lower halfspaces, respectively. The absorption in the nanoparticles was obtained using the density integration of the imaginary part of the electric field energy density inside the nanoparticles.

The material data used for the Ag nanoparticles was taken from Palik \cite{Palik:1991}, while a constant refractive index of 1.5 was assumed for glass.

\section{Results and Discussion}

\begin{figure}[h]
\centering
  \includegraphics[width=0.48\textwidth]{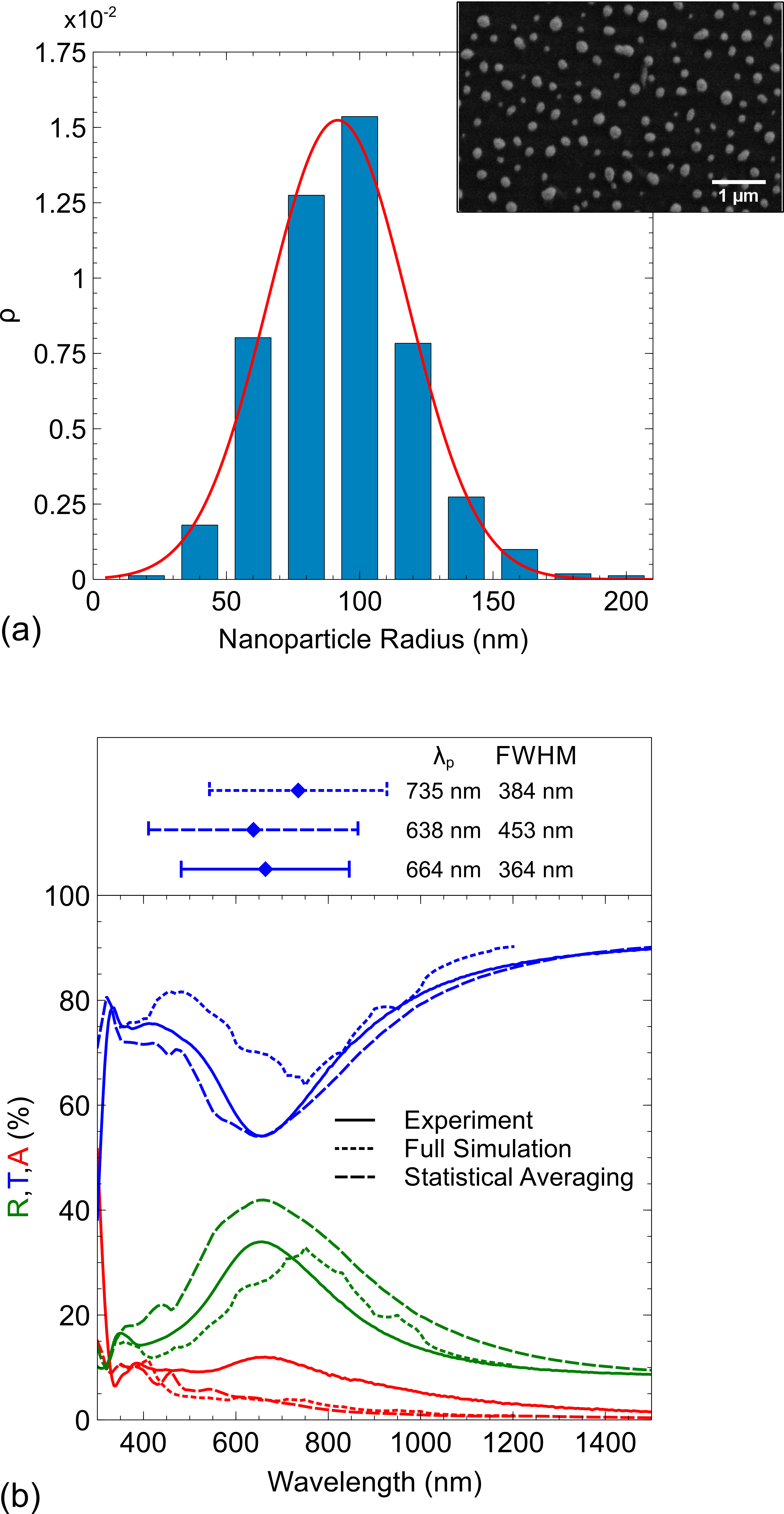}
  \caption{(a) The statistical distribution of particle radii determined from the SEM image of random nanoparticles shown in the inset. Histogram represents the measured statistical distribution, the red line gives a normal distribution fit to the histogram. (b) The transmittance (blue) reflectance (green) and absorption (red) for both experiment (solid lines) and simulation (dashed lines) of the nanoparticle distribution in part (a). The peak wavelength ($\lambda_{p}$) and FWHM of the transmittance resonance found through Gaussian fitting are shown above.}
  \label{fig:Disordered}
\end{figure} 	

Figure \ref{fig:Disordered}(a) shows the statistical distribution of radii taken from a sample comprising of a disordered array of Ag nanoparticles grown via thermal annealing. The particles are close to rotationally symmetric showing an average of 1.18 in the ratio of short to long axis. In order to further decrease the computational effort, we have modeled the particles as hemispheres with a radius given by the mean of the large and short axis taken from the SEM image. While this is not necessary for the statistical method, it allows the computational effort for simulations of the isolated nanoparticles to be greatly reduced by exploiting the rotational symmetry. The distribution of particle radii follows a normal distribution which is typical for the thermal annealing process \cite{Morawiec2013}, but it should be noted that other processes may give different statistical distributions such as log-normal. The distribution of radii is broad suggesting a wide range of particle sizes. Since plasmonic resonances are very sensitive to the lateral dimensions of particles, it is likely that the particle size will be the dominant factor in determining the resonance position. This is in contrast to the longitudinal dimension of the particle (w.r.t. the incident light) and the specific contact angle, which have been shown to shift the resonance to a lesser degree than the lateral dimensions \cite{Manley2015}. The overall coverage ($\rho$) of nanoparticles is 15\%, while the maximum extinction efficiency for hemispherical Ag particles on a glass substrate is 5.6 in the given radius range. This ensures the validation of the first condition since $0.15\times5.6 < 1$, meaning that on average no strong near field effects are expected.

Figure \ref{fig:Disordered}(b) shows the optical data for the distribution found in part (a). The solid lines represent the experimental result which is compared to two different simulation techniques. Firstly a direct simulation of the disordered array of particles with the same statistical distribution as that found in part (a) (dotted lines) and secondly the statistical averaging technique proposed in this paper (dashed lines). The grid used for the full particle simulation is shown in figure \ref{fig:Grids}(a).

To obtain the statistical response we averaged over 15 different particle radii, equally distributed in the range [25,200] nm, using the probability density function obtained from the experimental distribution.

The experimental data show a clear plasmonic resonance due to the particles, peaking at a wavelength of 664 nm. Light scattering from the nanoparticles increases the reflectance and absorption which both contribute to the drop in transmittance compared to bare glass. The full simulation of the nanoparticle array shows good agreement with the experimental reflectance except for a  redshift in the simulation result. This redshift is likely due to the fact that the statistical distribution of the particles included in the simulation could not perfectly match the statistical distribution of the experimental sample due to the finite number of particles in the simulation. This is a limitation which will always be present for full simulations of nanoparticle arrays unless very large domain sizes can be used which in turn increase the computational effort significantly. The transmittance and absorption show less good agreement with the experiment. The lower absorption present in the simulation is likely due to the neglection of light trapping in the glass substrate in the model. Light which is forward scattered by the particle may be trapped inside the glass substrate which would interact with the particles multiple times, meaning that the small absorption losses present in a single particle interaction are amplified. This leads to an increased transmittance and reduced absorption in the simulation compared to experiment.

For the simulated curves using the method of statistical averaging, the transmittance shows closer agreement with experiment than for the direct simulation, the peak position of the reflectance resonance is in better agreement with experiment, however the overall value of reflectance is overestimated compared to experiment. The absorption remains similar for both cases since they both neglect substrate light trapping. Obtaining the correct resonance wavelength for the simulated data is largely due to the ability to exactly reproduce the statistical parameters of the experimental nanoparticle distribution as will be demonstrated in the following section. The overestimation of the reflectance may be due to the assumed hemispherical shape of the particles.

It is important to note the computational effort required for both the statistical and full simulations. The full simulation required a median memory usage of 34 GB and a median CPU time of just over 3 hours per wavelength, calculated using 16 cores. In comparison the isolated single particle simulations used for the statistical method should be much less resource intensive. In the example presented here, the computational effort has been further reduced by exploiting the rotational symmetry of the particles. The median required memory usage for the isolated particles was 762 MB while the CPU time was 2 seconds per wavelength using a single core. This time should be multiplied by the number of different statistical parameters (in this case the nanoparticle radius) to average over. For the samples shown here, 15 different radii were used. Nevertheless, the amount of time used for the statistical method is still orders of magnitude less than for the full simulations.

\begin{figure}[h]
\centering
  \includegraphics[width=0.48\textwidth]{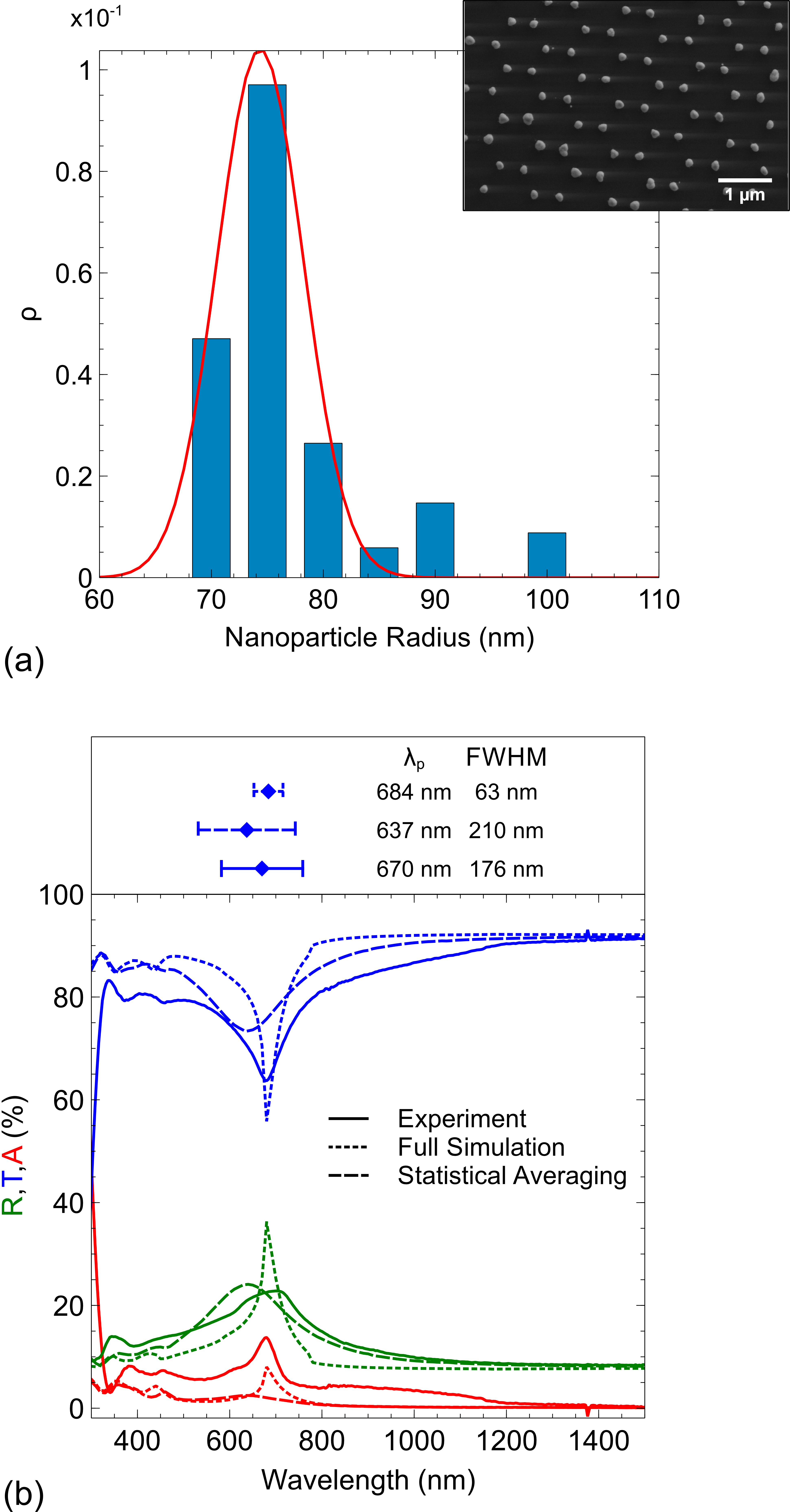}
  \caption{(a) The statistical distribution of particle radii determined from the SEM image of regular nanoparticles shown in the inset. Histogram represents the measured statistical distribution, the red line gives a normal distribution fit to the histogram. (b) The transmittance (blue) reflectance (green) and absorption (red) for both experiment (solid lines) and simulation (dashed lines) of the nanoparticle distribution in part (a). The peak wavelength ($\lambda_{p}$) and FWHM of the transmittance resonance found through Gaussian fitting are shown above.}
  \label{fig:Periodic}
\end{figure}

In order to quantify the agreement between the simulation and experiment Gaussian curves were fit to the dipolar resonance in the transmittance only. This was done since the reflectance and absorption resonances refer to different plasmon decay processes, namely scattering and absorption, respectively. The transmittance will be affected by both of these processes, thereby being able to fully characterise the plasmonic resonance regardless of decay process. The peak wavelength and full width at half maximum (FWHM) of the resonance can then be extracted from the Gaussian fit. Figure \ref{fig:Disordered}(b) shows the peak wavelength and FWHM of the Gaussian fit for transmittance from each data set. The relative error is defined for the peak wavelength as:
\begin{equation}
\Delta = \frac{|\lambda_{sim}-\lambda_{exp}|}{\lambda_{exp}}.
\label{equ:RelativeError}
\end{equation}
For the FWHM $\Delta$ is defined analogously. In this case the relative error for the peak wavelength is 0.11 for the full simulation and only 0.04 for the statistical averaging method. Conversely the relative error in the FWHM is 0.06 for the full simulation while increasing to 0.24 for the statistical averaging. Neither method is clearly superior to the other with regards to matching this single experimental result. However, it should be stressed that the method of statistical averaging can rapidly evaluate different statistical distributions, something which is not the case for the full simulation. This allows for a detailed scan of the statistical parameter space which is necessary for optimisation.

Figure \ref{fig:Periodic}(a) shows the statistical distribution of radii for an array of particles prepared using nanosphere lithography. The Ag nanoparticles are arranged in the honeycomb structure where a particle is present at each vertex of a hexagon due to the hexagonal close packed structure of the PS spheres (which have been removed). The distribution of radii is much narrower for the periodic sample which is to be expected since the nanosphere lithography technique seeks to obtain identical particles. Nevertheless some percentage of particles with a slightly larger radius lie outside the fitted normal distribution, however they are not expected to significantly contribute to the optical response and can be treated as outliers. The particle coverage in this case is 5\% which, as with the disordered array, means that near field effects can be eliminated. The second assumption present in the method of statistical averaging was a disordered array. It will now be shown that even for periodically ordered arrays, the results can still provide a good approximation to the particle resonance.

Figure \ref{fig:Periodic}(b) shows the optical response for the periodic array. The solid lines represent the experimental result which is compared to two different simulation approaches. Firstly a direct simulation of the honeycomb periodic array using periodic boundary conditions (dotted lines) and a particle radius taken from the peak of the statistical distribution (r = 75 nm), and secondly the statistical averaging technique proposed in this paper (dashed lines). Focusing first on the comparison between the experimental value and those from the method of statistical averaging, the resonance position is clearly blueshifted in the simulation compared to the experiment. This could possibly be due to an underestimation of the particle size, however the resonance position is more accurately reproduced using the periodic simulation, which is consistent with inter-particle interactions being the cause of the redshift. Since the nanoparticles are widely spaced no near field interactions are expected. Near field interactions would tend to split the plasmonic resonance into a higher and lower energy mode depending on the polarisation of the incident light compared to the axis joining the two particles, which is not observed. Since the particles have a fixed phase correlation due to their periodic ordering, the scattering will also be coherent in the far field, i.e. diffraction will occur. This is not the case for disordered arrays. The resonance position can be either red- or blue-shifted by this coherent superposition, and will tend to oscillate between the two depending on the inter-particle spacing. In the specific case presented here, the far field interaction between particles causes the redshift of the resonance compared to isolated particles of the same size.

The absorption is underestimated from both simulation methods compared to the experiment, this is again attributed to the effect of light trapping in the substrate. The peak in absorption at the plasmonic resonance (670 nm) which is present in the periodic simulation but absent in the statistical averaging simulation, suggests that the presence of this peak is also due to inter-particle interactions whereby part of the scattered field from each particle is absorbed by nearby neighbouring particles. The peak position and FWHM of a Gaussian fit to the dipolar resonance are shown above the optical response curves. Comparing the relative error in these two values compared to those extracted from the experimental data, it is observed that the relative error in the peak position is 0.02 and 0.06 for the full simulation and the statistical averaging, respectively. The FWHM shows a relative error of 0.64 and 0.19 for the full simulation and the statistical averaging, respectively. Therefore the peak position is in very good agreement for both methods, whereas the FWHM shows less good agreement for the full simulation. Additionally only the statistical averaging method is rapidly able to determine the optical response for different periodic nanoparticle size and coverage distributions.

Both particle distributions shown here were for low density particle distributions, having coverages of 15 and 5\% for the disordered and periodic arrays, respectively. This, coupled with the fact that the peak values of normalised scattering cross section for each isolated particle was relatively small (The largest being 5.6) means that the product of coverage and scattering cross section remains $<1$ for all wavelengths. As discussed previously, if the product of coverage and normalised scattering cross section exceeds unity then the statistical averaging model will break down. Therefore, since particle interactions are neglected, the method is restricted to low densities and/or particles which weakly scatter light.

\begin{figure}[h]
	\centering
	\includegraphics[width=0.48\textwidth]{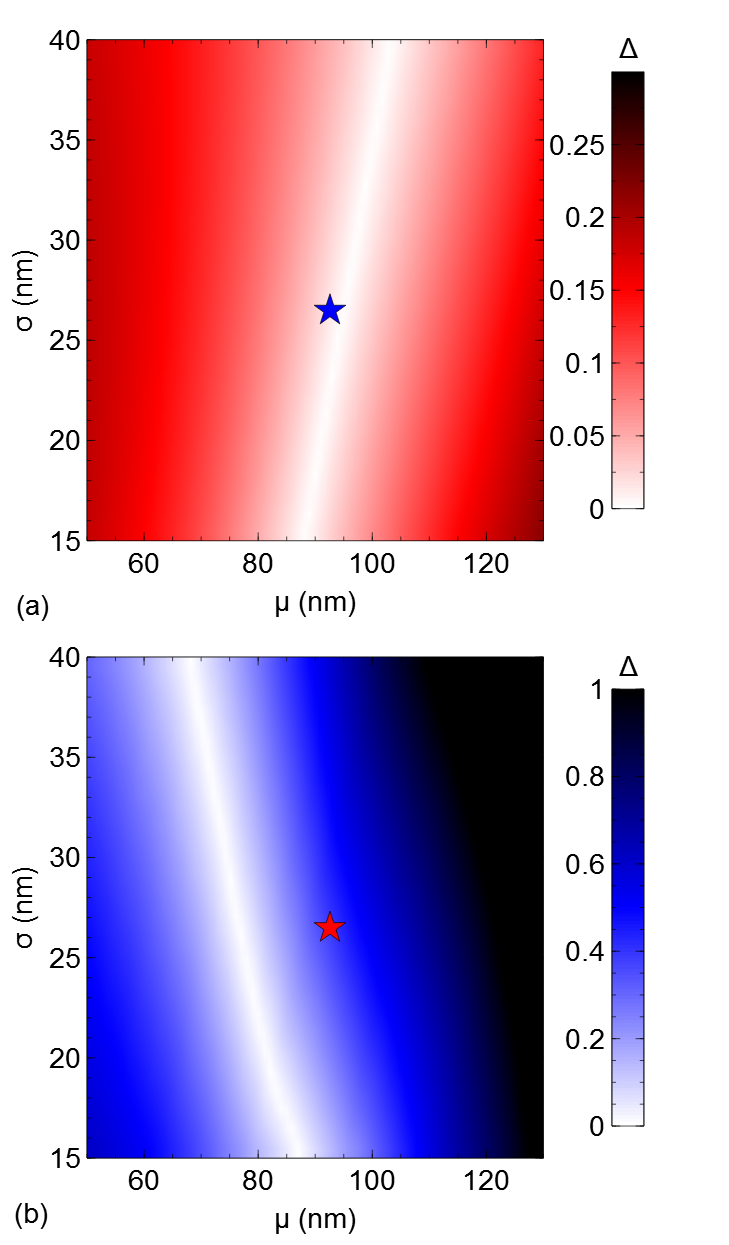}
	\caption{The residual error in resonance peak position (a) and resonance width (b) compared to the experimentally determined values for different assumed nanoparticle normal distribution mean values ($\mu$) and standard deviations ($\sigma$). In each case the experimentally determined mean and standard deviation are indicated by a star.}
	\label{fig:Optimisation}
\end{figure} 	

In figure \ref{fig:Optimisation} the specificity of the method is shown by computing the optical response of a wide range of normal distributions and comparing the relative error in resonance position (a) and FWHM (b) with respect to the experimental data shown in figure \ref{fig:Disordered} for each set of mean and standard deviation of the nanoparticle distribution. Due to the relative error being dependent on the statistical input parameters, only those parameters which correspond to the correct optical output will give a low relative error. This allows for the specification of statistical input parameters which will provide the desired optical response.

The peak position (figure \ref{fig:Optimisation}(a)) shows a simple response to the mean ($\mu$) and standard deviation ($\sigma$) of the particle radius. As the radius of nanoparticles increases, meaning larger $\mu$ values, the plasmonic resonance redshifts to larger wavelengths due to retardation effects \cite{Manley2015}. Therefore as the mean radius is increased, the resonance redshifts into the correct wavelength position to agree with experiment before being redshifted out of agreement with experiment. Note that in figure \ref{fig:Optimisation} it is not explicitly shown whether the simulated peak position is redshifted or blueshifted compared to the experimental result, only the absolute value of the relative error is shown as defined in equation \ref{equ:RelativeError}.

The effect of increasing the nanoparticle distribution width ($\sigma$) is to spread the statistical weighing more evenly over different nanoparticle sizes. For Ag nanoparticles with a radius less than 50 nm, scattering cross section increases with increasing radius, however for higher radius values, the peak value of scattering cross section tends to decrease due to radiative damping. By increasing $\sigma$ the statistical weighting of the different radii is more even. Therefore, in the case of even statistical weighting, the radii with a higher scattering will contribute more to the total scattering than those radii with a lower scattering. This culminates in the behaviour seen in figure \ref{fig:Optimisation}(a) where the increase in $\sigma$ increases the contribution to the total scattering from smaller nanoparticles (which have a higher scattering than larger ones for radii $> 50$ nm) which blueshifts the resonance. In order to be in agreement with the experimental data for the plasmonic peak wavelength, a higher mean value $\mu$ of the statistical distribution must be chosen to counteract this blueshift. This explains why the region of experimental agreement in figure \ref{fig:Optimisation}(a) shifts to high $\sigma$ values for higher $\mu$ values

The FWHM also shows a simple response to the statistical distribution parameters (figure \ref{fig:Optimisation}(b)); it is expected that when increasing the nanoparticle distribution standard deviation ($\sigma$), the resulting optical response will also be broadened. However when the mean particle radius becomes large, each individual resonance will also be broad, which means that a smaller standard deviation is required for the same optical broadness. This is shown by the decrease in standard deviation required to reach a particular FWHM with increasing mean nanoparticle radius. 

As can be seen from the star in parts (a) and (b), the experimentally determined statistical parameters give a solution in the optimal region for the peak position and close to the optimal region for the resonance width. Both cases quantify which regions of the statistical parameter space are capable of giving the correct optical resonance which is crucial for experimentally tuning the distributions. This is due to the fact that in many cases it is difficult to have complete control over the nanoparticle distribution and therefore the ability to define bounds on the required mean and standard deviation values for a given optical response is highly valuable.

The FWHM of the optical resonance was overestimated in the simulation as can be seen in figure \ref{fig:Disordered}, which, as previously discussed, is likely due to the neglection of light trapping effects in the glass substrate and difficulties in finding accurate values for the particle coverage.

\begin{figure}[h]
\centering
  \includegraphics[width=0.70\textwidth]{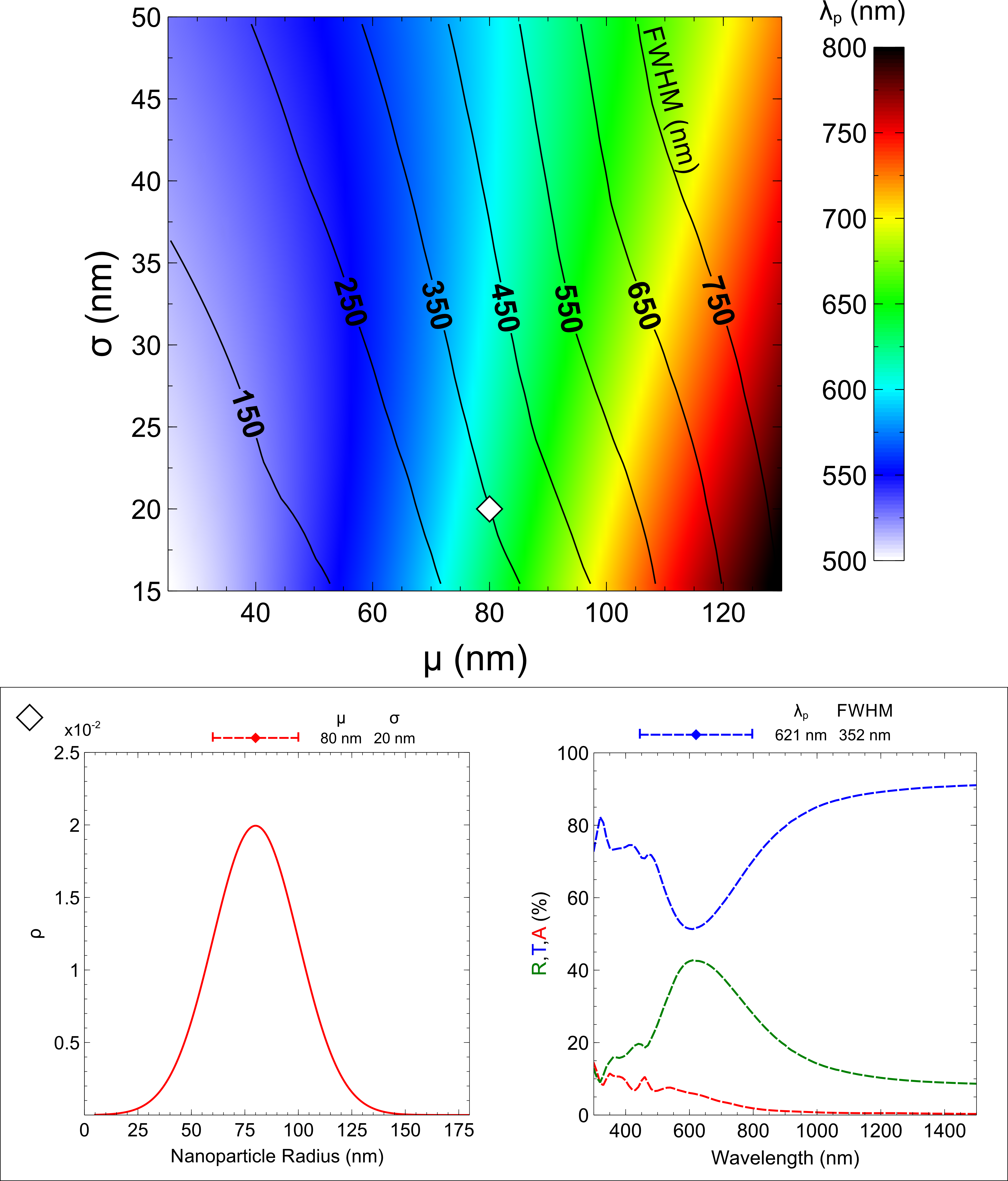}
  \caption{The optical resonance peak wavelength (colours) and full width at half maximum (contour lines) for different Ag nanoparticle distribution mean and standard deviation values. The box labeled with a diamond shows for reference the nanoparticle distribution and optical response of the distribution shown by the diamond in the contour image.}
  \label{fig:LookUpTable}
\end{figure}

It should be emphasised that such an analysis of different statistical parameters is easy to generate once the individual particle responses have been calculated. On the other hand, direct simulation of the nanoparticle distribution requires an entire spectrum calculation for each new mean and standard deviation of the nanoparticle distribution meaning that many orders of magnitude of extra computational effort will be required for a similar analysis.

In figure \ref{fig:LookUpTable} the resulting optical peak wavelength and FWHM are shown for a wide range of input nanoparticle mean and standard deviations (normal distribution) of the particle radii of Ag plasmonic nanoparticles on a glass substrate. The colour scale refers to the resonance peak wavelength while the contour lines refer to the FWHM of the resonance. An example of the nanoparticle statistical distribution and the resulting optical curves is given in the box marked for a specific example of input mean and standard deviation. Above the statistical distribution the mean $\mu$ and standard deviation $\sigma$ are shown, similarly the peak wavelength ($\lambda_{p}$) and FWHM of the optical resonance found from a Gaussian fitting are also shown. This can be used to predict the optical resonance of Ag nanoparticle distributions on a glass substrate. For other substrates of higher refractive index than glass ($n$ = 1.5) the peak position will be redshifted, for a 100 nm radius particle the redshift in substrate index causes a linear shift of $\lambda_{p}(n) = n \times 450$~nm$ + 41$~nm, where $n$ is the substrate refractive index. This can be used to estimate the peak position for other substrate materials. For a more accurate determination for other substrate materials and for different nanoparticle shapes (e.g. cylinders, triangles) or materials (e.g. Au, Al) the method presented here can be used to quickly determine the optical response of an array based on the single particle responses. 

\begin{table}
	\renewcommand\arraystretch{1.3}
		\centering
	\begin{tabular}{|c|c|c|c|c|c|}
		\hline		
		Ref.& $\mu$ (nm) & $\sigma$ (nm) & $\rho$ &$\lambda_{p}$ (nm)&  FWHM (nm)\\
		\hline
		\cite{Schmid2011} & 30 & 10 & 8\%& 432 (420) & 92 (159) \\
		\hline
		\cite{Santbergen2012} & 9 & 3 & 34\%& 411 (440) &59 (166) \\
		\hline
		\cite{Thouti2013} & 75 & 15 & 24\%& 505 (458) &280 (126)\\
		\hline
		\cite{Choi2007} & 50 & 25 & 24\%& 459 (525) &167 (151)\\
		\hline
		\cite{Giallongo2013} & 12 & 2.5 & 10\%& 412 (407) & 50 (83)\\
		\hline
		
	\end{tabular}
\caption{The Ag nanoparticle size distribution parameters ($\mu$, $\sigma$ and $\rho$) and the simulated  optical resonance parameters ($\lambda_{p}$, FWHM). In parentheses are the experimentally determined values for the optical response presented in the same reference.}	
\label{tab:LitComparison}
\end{table}

In order to further test the limits of the method, we have simulated the optical response for five different nanoparticle distributions taken from literature, and compared them with their experimentally determined values. The optical response parameters and their respective nanoparticle distribution parameters are given in table \ref{tab:LitComparison}. We have assumed a truncated spherical particle shape with the total height of the particle equal to 90\% of the diameter in order to be in agreement with the nanoparticle shapes reported in \cite{Schmid2011,Santbergen2012,Thouti2013,Choi2007,Giallongo2013}. The size distribution parameters ($\mu$ and $\sigma$) were obtained directly from the paper when present \cite{Thouti2013,Giallongo2013}, or by analysis of a suitable SEM image present in the work \cite{Schmid2011,Santbergen2012} or estimated from a smaller SEM image \cite{Choi2007}. Nanoparticle coverage ($\rho$) was obtained directly from the text \cite{Schmid2011,Giallongo2013} or obtained via image analysis of a SEM image \cite{Santbergen2012,Thouti2013,Choi2007}.

Our method is able to accurately predict the peak resonance wavelength ($\lambda_p$) for low particle coverages ($\rho$) while showing larger deviations for higher particle coverage values. This agrees with the assumption that neglecting particle interactions is only valid for low particle densities. The full width half maximum (FWHM) shows less agreement with the experimental data. The main factor affecting the FWHM determined via our method is the standard deviation of particle sizes ($\sigma$). This was often difficult to obtain from the literature as it is not as commonly reported as the mean particle size. In addition to this, other factors such as the non-spherical shape, contact angle and particle interactions certainly also affect the resonance width. While it is possible to take the shape and contact angle factors into account using our method, it is challenging to characterise the nanoparticles experimentally to obtain the required information. 

\FloatBarrier
\section{Conclusion}
We present a method for quickly calculating the optical response of an array of nanoparticles. This can be used to efficiently search the design space of nanoparticle distributions required for a certain application, thereby solving the design problem. The method relies on treating different particle sizes separately and determining the average optical response from the single particle responses and the statistical distribution of particle sizes. We compared the optical response of two samples, a disordered and an ordered array, to their simulated optical responses. Our method was found to be in good agreement with the measured optical data. For the case of the ordered array we observed a blueshift compared to experiment due to neglection of inter-particle interactions. By calculating the optical response of artificial nanoparticle distributions, we showed the specificity of the method. This revealed that a distinct zone of mean and standard deviations for the normal distribution of particle sizes leads to agreement with the experimental optical resonance. This result is important experimentally as it allows for a margin of error in preparing the nanoparticle distributions while still maintaining the correct optical response. Finally we provide a look-up graph as a reference for determining the correct nanoparticle distribution of Ag nanoparticles on glass for a desired optical resonance. This table can easily be extended to other substrate and particle materials, and particle shapes and sizes using the presented method. Further work should compare the optical results to more experimental data in order to assess the limits of applicability of the model.

\section{Acknowledgments}
The authors would like to thank D. Lockau and M. Hammerschmidt for their help in the simulation of multiple nanoparticles, F. Schmidt for discussion on simulation of multiple particles and P. Andr\"{a}  and G. Yin for fruitful advice on the preparation of nanoparticle samples.

P. Manley, M. Song and M. Schmid would like to acknowledge funding and support from the Initiative and Networking fund of the Helmholtz Association for the Young Investigator Group VH-NG-928. Part of the work was done at the Berlin Joint Lab for Optical Simulations for Energy Research (BerOSE). M. Song acknowledges the support of funding from the China Scholarship Council. P. Manley acknowledges funding from the Helmholtz Innovation Lab HySPRINT, which is financially supported by the Helmholtz Association. 


\vspace{1cm}




\bibliography{Bibliography} 
\bibliographystyle{unsrt} 

\end{document}